\documentclass[
intlimits,
sumlimits,
namelimits,
10pt,
a4paper,
]{article}

\usepackage{arxiv}

\theoremseparator{:}%

\usepackage{xifthen}

\usepackage{cmap}
\usepackage{textcomp}
\usepackage{mathtext}

\usepackage[T1]{fontenc}
\usepackage[utf8]{inputenc}%
\usepackage[main=english]{babel}

\usepackage{microtype}      % microtypography
\UseMicrotypeSet[protrusion]{alltext}

\ifx\No\undefined
    \def\No{\textnumero}
\fi

\usepackage{etex}

\usepackage{bm}
\usepackage{epigraph}
\usepackage{layout}

\usepackage{lscape}
\usepackage{epic}

\usepackage{soulutf8}

\usepackage{texnames}
\usepackage{paralist}

\usepackage{rotating}

\usepackage{wrapfig}

\usepackage[shortcuts]{extdash}

\usepackage{setspace}
\usepackage[perpage,bottom,multiple,stable]{footmisc}

\RequirePackage{color}
\usepackage{hyperref}
\hypersetup{backref,
linktoc=all
}
\hypersetup{pdfborder=0 0 0}

\hypersetup{pdfencoding=auto}

\ifthenelse{\boolean{xetex}}{}{%
  \usepackage{breakurl}
}

\definecolor{darkred}{rgb}{.5,0,0}
\definecolor{darkgoldenrod}{rgb}{.7,.5,0}
\definecolor{darkorange}{rgb}{.7,.4,0}
\definecolor{darkteal}{rgb}{0,.2,.2}
\definecolor{darkmagenta}{rgb}{.5,0,.5}
\definecolor{darkyellow}{rgb}{.5,.5,0}
\definecolor{darkgreen}{rgb}{0,.5,0}
\definecolor{darkblue}{rgb}{0,0,.5}
\definecolor{darkcyan}{rgb}{0,.5,.5}

\hypersetup{baseurl=http://, unicode, colorlinks, linkcolor=darkred, filecolor=darkteal, citecolor=darkgoldenrod, urlcolor=darkmagenta}

\usepackage{amsmath}
\usepackage{amssymb}
\usepackage{amsfonts}       % blackboard math symbols
\usepackage{nicefrac}       % compact symbols for 1/2, etc.
\usepackage{amscd}
\usepackage{slashed} % для обозначений по Фейнману "/"

\usepackage{accents}

\usepackage{mathtools}
\mathtoolsset{
showonlyrefs=true,
}

\allowdisplaybreaks

\RequirePackage{color}

\usepackage{pgfplotstable}
\pgfplotsset{compat=1.8}
\usepackage{booktabs}       % professional-quality tables
\usepackage{array}
\usepackage{colortbl}
\usepackage{multirow}

\usepackage{graphicx}

   \usepackage{float}
\graphicspath{{image/}}

\usepackage[scanall]{psfrag}

\usepackage{tikz}
\usetikzlibrary{arrows,shapes,graphs}

\usepackage{listingsutf8}
\ifthenelse{\boolean{luatex}\OR\boolean{xetex}}{}{%
   \lstset{inputencoding=utf8/koi8-r}}

\usepackage{calc}

\usepackage{url}

 \usepackage{cite}

\ifdefined\refname\ifx\refname\empty\renewcommand*{\refname}{Литература}\else\renewcommand{\refname}{Литература}\fi\else\newcommand*{\refname}{Литература}\fi
\ifdefined\abstractname{\renewcommand{\abstractname}{}}\fi

\ifdefined\sectionFontShape{\renewcommand{\sectionFontShape}{\bfseries}}\else{\newcommand{\sectionFontShape}{\bfseries}}\fi
\ifdefined\subsectionFontShape{\renewcommand{\subsectionFontShape}{\bfseries}}\else{\newcommand{\subsectionFontShape}{\bfseries}}\fi
\ifdefined\subsubsectionFontShape{\renewcommand{\subsubsectionFontShape}{\bfseries}}\else{\newcommand{\subsubsectionFontShape}{\bfseries}}\fi
\ifdefined\paragraphFontShape{\renewcommand{\paragraphFontShape}{\bfseries}}\else{\newcommand{\paragraphFontShape}{\bfseries}}\fi
\ifdefined\subparagraphFontShape{\renewcommand{\subparagraphFontShape}{\bfseries}}\else{\newcommand{\subparagraphFontShape}{\bfseries}}\fi
\ifdefined\fi
\begin{document}

	{\selectlanguage{english}

		%% Подключение файла с титульным листом
		% титульная часть

% \crop[cam,noinfo,axes]

% Универсальный десятичный классификатор
% \udc{537.8}

\title[The Einstein photon and Dirac electron-positron fields]{%
   On certain paradoxical properties of Einstein photon fields\\
   and Dirac electron-positron fields
}

% \alttitle{%
   % Обобщённая квантовая механика фотона Эйнштейна\\
   % и сила Казимира
% }

% \def\address{%
%    Department of Theoretical Physics \\
%    Russian Peoples' Friendship University \\
%    Russia, 117198, Moscow, Miklukho-Maklaya st., 6
% }
% \def\authorone{\href{https://orcid.org/0000-0003-4474-9016}{\includegraphics[scale=0.06]{orcid.eps}\hspace{1mm}Beilinson A.~A.}}

\author[a]{Beilinson A.~A.}
% \author{ \href{https://orcid.org/0000-0003-4474-9016}{\includegraphics[scale=0.06]{orcid.eps}\hspace{1mm}Beilinson A.~A.} \\
% 	\address \\
% 	\texttt{alalbeyl@gmail.com} \\
% }

% \altauthor[a]{Бейлинсон А. А.}%

\address[a]{
   Department of Theoretical Physics \\
   Russian Peoples' Friendship University \\
   Russia, 117198, Moscow, Miklukho-Maklaya st., 6
}

% \altaddress[a]{%
%     Кафедра теоретической физики \\
%     Российский университет дружбы народов \\
%     Россия, 117198, Москва, ул. Миклухо-Маклая, 6
% }

\email[work]{alalbeyl@gmail.com}

% \institute[PFUR]{Russian Peoples' Friendship University}

% \altinstitute[РУДН]{Российский университет дружбы народов}

% \thanks{%
  %
% }

\begin{abstract}
   \indent The article shows that the photons and electrons in the states with {\guillemotleft}deinterlaced{\guillemotright} spins, which are isomorphic to their usual wave states, turned out to be not only nonlocal states on {\guillemotleft}Feynman paths{\guillemotright}, but also having a power to penetrate infinitely distant in arbitrarily small time.
\end{abstract}

% \begin{altabstract}
%    \indent В статье показывается, что фотоны и электроны в состоянииях с {\guillemotleft}расплетенными{\guillemotright} спинами, являющимися изоморфными их обычным волновым состояниям, оказались не только нелокальными состояниями на {\guillemotleft}путях Фейнмана{\guillemotright}, но и обладающими способностью проникать бесконечно далеко за сколь угодно малое время.
% \end{altabstract}


% \altkeywords{%
  % обобщённая квантовая механика, обощённый интеграл по путям, Сила Казимира, пути Фейнмана, квантовое поле фотона.
% }

% \subject{Квантовая механика}

% \received{\formatdate{31}{12}{1917}}

% Here you can change the date presented in the paper title
%\date{September 9, 1985}
% Or remove it
\date{}

% Uncomment to override  the `A preprint' in the header
% \renewcommand{\headeright}{}
% \renewcommand{\undertitle}{}
% \def\keywordname{{\bfseries {Keywords and phrases}}}%
% \renewcommand{\shorttitle}{The Einstein photon and Casimir force}

%%% Add PDF metadata to help others organize their library
%%% Once the PDF is generated, you can check the metadata with
%%% $ pdfinfo template.pdf
\hypersetup{%
pdftitle={On certain paradoxical properties of Einstein photon fields and Dirac electron-positron fields},
pdfsubject={math-ph, quant-pm},
pdfauthor={Beilinson A.~A.},
pdfkeywords={generalized quantum mechanics, generalized paths integral, Casimir force, Feynman paths, quantum field of a photon},
}

\renewcommand{\headrulewidth}{0pt}
\fancyheadoffset{0pt}
\lhead{}
\rhead{}
\chead{}
\lfoot{}
\rfoot{}
\cfoot{\thepage}
\copyrightinfo{}{}

\maketitle

%%% Local Variables:
%%% mode: latex
%%% coding: utf-8-unix
%%% TeX-master: "./default"
%%% End: %

		%% Подключение файла с аннотацией
		%% аннотация

% {\small

% }

% \indent {\bfseries Ключевые слова:}

%%% Local Variables:
%%% mode: latex
%%% coding: utf-8-unix
%%% TeX-master: "../default"
%%% End:%

		%% Подключение файла с оглавлением
		%% оглавление

% \tableofcontents

%%% Local Variables:
%%% mode: latex
%%% coding: utf-8-unix
%%% TeX-master: "../default"
%%% End:%

		%% Подключение файла с введением
		%% введение
\section*{Introduction}
% % \addtocontents{toc}{section}{Введение}
\label{sec:introduction}

\indent This refers to a number of reliable experimental data, according to which the speed of propagation of signals generated by «entangled» photons and electrons exceeds many times the speed of light,~see for example~\cite{nobelprize:physics:2022:01}.

\indent The article shows that in the scope of generalized solutions of equations for an Einstein photon e.\nobreak-m. field and Dirac electron-positron field these physical quantum fields along with existance of nonlocal states of corresponding relativistic quantum particles and their path integrals, and mentioned paradoxical properties, also can be sucessfully described,~see~\cite{bejlinson:article:inproc:2018:01, bejlinson:article:inproc:2018:02, bejlinson:article:2023:01}.

\indent The article shows that these phenomena arises due to unitarily equivalent states of relativistic quantum particles associated with cut-off their spin interactions. A similar phenomenon in relativistic quantum particle physics has actually already been considered using known Foldy\nobreak--Wouthuysen variables,~see~\cite{foldy:article:1950:01}.

\indent For simplicity onedimensional fields will be considered. The speed of light assuming equal~$1$; More detailed the case of e.\nobreak-m. field is considered.

The work uses notation and terminology corresponding to monographs «Generalized functions», issues 1, 2 and 4 by I.~M.~Gel'fand, N.~Ya.~Vilenkin and G.~E.~Shilov~(see~\cite{gelfand:books:02:iss:01, gelfand:books:02:iss:02, gelfand:books:02:iss:04}).

%%% Local Variables:
%%% mode: latex
%%% coding: utf-8-unix
%%% TeX-master: "../default"
%%% End:%

		%% Подключение файла с основным содержанием
		%% основное содержание
% Обобщённая функция Грина одномерного квантового э.-м. поля как функционал на финитных функциях
\section*{Two record possibilities for generalized retarded Green functional of e.\nobreak-m. field and Dirac field}
\label{sec:two_record_possibilities_for_generalized_retarded_green_functional_of_em_field_and_dirac_field}

\indent Recall that Maxwell equations of e.\nobreak-m. field in Majorana variables~\cite{axiezer:book:1981:01}
\begin{equation}
	\label{eq:majorana_variables}
	\begin{aligned}
		& M_{t}(x) = E_{t}(x) + i H_{t}(x) , & \bar{M}_{t}(x) = E_{t}(x) - i H_{t}(x) ,
	\end{aligned}
\end{equation}
assuming this field is fundamentally quantum~(see~\cite{axiezer:book:1981:01, bejlinson:article:2023:01}), are becomes
\begin{equation}
	\label{eq:maxwell_equations_in_majorana_variables}
	\begin{aligned}
		& i \frac{\partial}{\partial{t}} M_{t}(x) = (S,\hat{p}) M_{t}(x) , & -i \frac{\partial}{\partial{t}} \bar{M}_{t}(x) = (S,\hat{p}) \bar{M}_{t}(x) ,
	\end{aligned}
\end{equation}
where $\hat{p} = \frac{1}{i} \nabla$, and $S$ is spin operators of a photon~$M_{t}(x)$,~see~\cite{axiezer:book:1981:01}); antiphoton~$bar{M}_{t}(x)$ is considered similarly.

\indent In momentum representation these equations are as follows:
\begin{equation}
	\label{eq:impulse_maxwell_equations_in_majorana_variables}
	\begin{aligned}
		& i \frac{\partial}{\partial{t}} \tilde{M}_{t}(p) = (S,p) \tilde{M}_{t}(p) , & -i \frac{\partial}{\partial{t}} \bar{\tilde{M}}_{t}(p) = (S,p) \bar{\tilde{M}}_{t}(p) .
	\end{aligned}
\end{equation}
We will consider such motion of a photon along the $z$ axis, in which the coordinates $x$, $y$ is any. To this e.\nobreak-m. field, it is easy to see, corresponding to it the three of the momentum representations of it wave functions
\begin{equation}
	\label{eq:impulse_wave_functions_for_onedimensional_photon_motion}
	\begin{aligned}
		& \tilde{M}_{t}(p_{z}) =
		\begin{pmatrix*}[r]
			\tilde{M}_{t}^{(1)}(p_{z}) \\
			\tilde{M}_{t}^{(2)}(p_{z})
		\end{pmatrix*}
		,
		& \tilde{M}_{t}^{(3)}(p_{z}) ,
	\end{aligned}
\end{equation}
satisfying the equations
\begin{equation}
	\label{eq:impulse_equations_for_onedimensional_photon_motion}
	\begin{aligned}
		& i \frac{\partial}{\partial{t}} \tilde{M}_{t}(p_{z}) =
		\begin{pmatrix*}[r]
			0 ,& -ip_{z} \\
			ip_{z} ,& 0
		\end{pmatrix*}
		\tilde{M}_{t}(p_{z}) ,
		& i \frac{\partial}{\partial{t}} \tilde{M}_{t}^{(3)}(p_{z}) = 0 ,
	\end{aligned}
\end{equation}
whose retarded Green function in momentum representation is
\begin{equation}
	\label{eq:impulse_green_function_for_onedimensional_photon_motion}
%   \begin{gathered}
      \begin{aligned}
         & \tilde{M}_{t}(p_{z}) = \exp\!\left( -it
         \begin{pmatrix*}[r]
            0 ,& -ip_{z} \\
            ip_{z} ,& 0
         \end{pmatrix*}
         \right) = \exp\!\left( -i
         \begin{pmatrix*}[r]
            0 ,& -itp_{z} \\
            itp_{z} ,& 0
         \end{pmatrix*}
         \right) ,
         & \tilde{M}_{t}^{(3)}(p_{z}) = 1 .
      \end{aligned}
%\end{gathered}
\end{equation}
(we will not consider last equation).

\indent Bearing in mind the constructing of the coordinate representation the generalized Green function of this field as functional on columns of bump functions $\varphi(z) \in K$ (as Fourier preimage of solution in the momentum representation,~see~\cite{gelfand:books:02:iss:02}), we will first consider the latter as a functional
\begin{equation}
	\label{eq:solution_of_impulse_equations_for_onedimensional_photon_motion}
	\int \bar{\tilde{M}}_{t}(p_{z})
   \begin{pmatrix*}[r]
      \psi_{1}(p_{z}) \\
      \psi_{2}(p_{z})
   \end{pmatrix*}
   \mathop{}\!{d{p_{z}}}
\end{equation}
on columns of analytic test functions $\psi(p) \in Z$~(see~\cite{bejlinson:article:2023:01}).

\indent Remark that the momentum representation of solution contains a Hermitian matrix in the exponent, which can therefore be reduced to a diagonal form by unitary transform $\tilde{Q}(p_{z})$. Therefore we have
\begin{equation}
	\label{eq:impulse_green_function_for_onedimensional_photon_motion:diagonalization}
   \tilde{M}_{t}(p_{z}) = \exp\!\left( -i
   \begin{pmatrix*}[r]
      0 ,& -itp_{z} \\
      itp_{z} ,& 0
   \end{pmatrix*}
   \right) = \tilde{Q}^{+}(p_{z})
   \begin{pmatrix*}[r]
      \exp(-it|p_{z}|) ,& 0 \\
      0 ,& \exp(it|p_{z}|)
   \end{pmatrix*}
   \tilde{Q}(p_{z}) ,
\end{equation}
where
\begin{equation}
	\label{eq:impulse_foldy_wouthuysen_operator_for_onedimensional_photon_motion_green_function}
   \tilde{Q}(p_{z}) = \frac{1}{\sqrt{2}}
   \begin{pmatrix*}[r]
      -i \operatorname{sgn} p_{z} ,& 1 \\
      i \operatorname{sgn} p_{z} ,& 1
   \end{pmatrix*}
   .
\end{equation}
Note that arised diagonal generalized Green function can be called the Green function of e.\nobreak-m. field in Foldy\nobreak--Wouthuysen variables~(see~\cite{foldy:article:1950:01}).

Therefore, it is easy to see,
\begin{equation}
	\label{eq:impulse_green_function_for_onedimensional_photon_motion:other_view}
   %\begin{gathered}
      \tilde{M}_{t}(p_{z}) = \frac{1}{2}
      \begin{pmatrix*}[r]
         \exp(-it|p_{z}|) + \exp(it|p_{z}|) ,& \operatorname{sgn}(p_{z}) \cdot 2i \sin(t|p_{z}|) \\
         -\operatorname{sgn}(p_{z}) \cdot 2i \sin(t|p_{z}|) ,& \exp(it|p_{z}|) + \exp(-it|p_{z}|)
      \end{pmatrix*}
     % = \\
      =
      \begin{pmatrix*}[r]
         \cos(tp_{z}) ,& \sin(tp_{z}) \\
         -\sin(tp_{z}) ,& \cos(tp_{z})
      \end{pmatrix*}
      .
   %\end{gathered}
\end{equation}
But the Fourier preimages of numerical functionals $\exp(it|p_{z}|)$ and $\operatorname{sgn} p_{z}$ on $Z$ are the quantum Cauchy functional~(see~\cite{bejlinson:article:inproc:2018:01})
\begin{equation}
	\label{eq:onedimensional_quantum_cauchy_functional}
   C_{it}(z) = \frac{1}{2} \left( \delta(t - z) + \delta(t + z) \right) + \frac{i}{2 \pi} \left( \frac{1}{t - z} + \frac{1}{t + z} \right)
\end{equation}
and, accordingly, the functional $-\frac{i}{\pi z}$~(see~\cite{gelfand:books:02:iss:01},~p.~360~formula~19) on bump functions $\varphi(z) \in K$.

\indent Hence it is easy to see that we have coordinate representation the generalized Green function
\begin{equation}
	\label{eq:solution_of_equations_for_onedimensional_photon_motion}
	\int \bar{M}_{t}
   \begin{pmatrix*}[r]
      \varphi_{1} \\
      \varphi_{2}
   \end{pmatrix*}
   \mathop{}\!{d{z}}
\end{equation}
of Maxwell\nobreak--Majorana equations, where
\begin{equation}
	\label{eq:green_function_for_onedimensional_photon_motion:other_view}
   M_{t}(z) = \frac{1}{2}
   \begin{pmatrix*}[r]
      \delta(t - z) + \delta(t + z),& i \left( \delta(t - z) - \delta(t + z) \right) \\
      -i \left( \delta(t - z) - \delta(t + z) \right),& \delta(t - z) + \delta(t + z)
   \end{pmatrix*}
   .
\end{equation}
Remark that conducted diagonalization operation of the retarded Green function of e.\nobreak-m. field have physical interpretation, since diagonalization of matrix underlying of this operation is interpreted as cut-off the spin interaction between components of the considered generalized e.\nobreak-m. field of a photon and appearance of {\guillemotleft}new{\guillemotright} field in the result with the Green function momentum representation
\begin{equation}
	\label{eq:impulse_green_function_for_onedimensional_quantum_cauchy_field}
   \tilde{m}_{t}(p_{z}) =
   \begin{pmatrix*}[r]
      \exp(-it|p_{z}|) ,& 0 \\
      0 ,& \exp(it|p_{z}|)
   \end{pmatrix*}
   = \exp\!\left( it
   \begin{pmatrix*}[r]
      -|p_{z}| ,& 0 \\
      0 ,& |p_{z}|
   \end{pmatrix*}
   \right) ,
\end{equation}
with {\guillemotleft}deinterlaced{\guillemotright} components. Here the unitary operator $\tilde{Q}(p_{z})$ realizes both {\guillemotleft}deinterlacing{\guillemotright} and {\guillemotleft}interlacing{\guillemotright} of the filed components, and their Fourier preimage as functionals on bump functions
\begin{equation}
	\label{eq:foldy_wouthuysen_operator_for_onedimensional_cauchy_field_green_function}
   \begin{aligned}
      & Q(z) = \frac{1}{\sqrt{2}}
      \begin{pmatrix*}[r]
         \frac{-1}{\pi z} ,& \delta(z) \\
         \frac{1}{\pi z} ,& \delta(z)
      \end{pmatrix*}
      ,
      & Q^{+}(z) \mathrel{\ast} Q(z) =
      \begin{pmatrix*}[r]
         \delta(z) ,& 0 \\
         0 ,& \delta(z)
      \end{pmatrix*}
      .
   \end{aligned}
\end{equation}
realizes {\guillemotleft}interlacing{\guillemotright} and {\guillemotleft}deinterlacing{\guillemotright} in the coordinate representation.

\indent Here the Green function coordinate representation as functional on bump functions, it is easy to see, is
\begin{equation}
	\label{eq:green_function_for_onedimensional_photon_motion}
	M_{t}(z) = Q^{+}(z) \mathrel{\ast} m_{t}(z) \mathrel{\ast} Q(z)
\end{equation}

\indent Compare one more $M_{t}(z)$ and $m_{t}(z)$. These generalized retarded Green functions of e.\nobreak-m. field of a photon are unitarily equivalent and isomorphic; despite of that they are represented of one initial state development, this is radially different branch of a photon quantum state development: $M_{t}(z)$ is wave one propagating with fundamental velocity, and other one is $m_{t}(z)$ filling whole coordinate space in arbitrarily small time and, as has been shown, has nonlocal generalized states on {\guillemotleft}Feynman paths{\guillemotright}~(see~\cite{bejlinson:article:inproc:2018:02, bejlinson:article:2023:01}.

\indent Thus the isomorphism of {\guillemotleft}deinterlaced{\guillemotright} and wave states indicate the existance of integral over {\guillemotleft}Feynman paths{\guillemotright} also in case of wave brach solution $M_{t}(z)$.

\indent Let us consider case of Dirac electron-positron field, the retarded Green function of which is delivered by Pauli\nobreak--Jordan formula that also is functionals on bump functions. As has been shown, the diagonalization of the Hamiltonian of this functional reducing to known Foldy\nobreak--Wouthuysen variables~(see~\cite{foldy:article:1950:01}) also is interpreted as {\guillemotleft}deinterlacing{\guillemotright} of the wave functional components when disabling spin interaction and reduces to the Green functional of Dirac electron~(see~\cite{bejlinson:article:inproc:2018:01})
\begin{equation}
  %\label{eq:onedimensional_dirac_field_green_function:foldy_wouthuysen_variables}
  \label{eq:onedimensional_foldy_wouthuysen_field_green_function}
  \begin{pmatrix*}[r]
    C_{it}^{m}(z) ,& 0 \\
    0 ,& C_{it}^{m}(z)
  \end{pmatrix*}
%\end{equation}
  , \text{where}\
%\begin{equation}
 % \label{eq:onedimensional_foldy_wouthuysen_field_green_function}
  C_{it}^{m}(z) = \frac{1}{2} \left( \delta(t - z) + \delta(t + z) \right) + \frac{1}{\pi} \cdot \frac{-tm K_{1}\!\left( im\sqrt{t^{2} - z^{2}} \right)}{\sqrt{t^{2} - z^{2}}} .
\end{equation}
Thus also in case of Dirac field arising the unitarily equivalent isomorphic form of the field solutions that has lost its usual wave properties and acquired a property instant propogation in whole sapce. It can be shown that, together with this property, this form acquired exactly the same properties of nonlocality as the photon field. Here the corresponding generalized quantum measures are turned out to be absolutely continuous.

It should be remarked that all the above constructions, it is easy to see, are Lorentz-covariant.

%%% Local Variables:
%%% mode: latex
%%% coding: utf-8-unix
%%% TeX-master: "./default"
%%% End:
%

		%% Подключение файла с заключением
		%% заключение
%\section*{Conclusion}
%\label{sec:conclusion}

%%% Local Variables:
%%% mode: latex
%%% coding: utf-8-unix
%%% TeX-master: "../default"
%%% End:%

		%% Подключение файла со списком публикаций
		%% список литературы

% \section*{\refname}
% \addtocontents{toc}{section}{Список литературы}
\label{sec:references}

% \bibliographystyle{ugost2008ls}
% \putbib[bib/alt]
 \bibliography{default}
%\input default.bbl

%%% Local Variables:
%%% mode: latex
%%% coding: utf-8-unix
%%% TeX-master: "../default"
%%% End:%

		%% Подключение файла с приложением
		% приложение
% \section{Приложение}
% \label{sec:application}

%%% Local Variables:
%%% mode: latex
%%% coding: utf-8-unix
%%% TeX-master: "./default"
%%% End:%

	} % END \selectlanguage

\end{document}